# Trends and patterns of scintillator nonproportionality

Ivan V. Khodyuk and Pieter Dorenbos

*Abstract*— Data on the photon nonproportional response of 33 inorganic scintillation materials are systemized and analyzed. The main trends of nonproportionality for different groups of inorganic scintillators, especially for oxides and halides, are highlighted. The dependence of the shape and degree of photon nonproportional response versus chemical composition, dopant type, refractive index and other fundamental properties of the materials is studied. Better proportionality appears to be correlated with higher refractive index of the compound. Another related factor is the width of the valence band in halide compounds. With larger valence band width from fluorides, to chlorides, to bromides, and to iodides, a better proportionality is observed.

## I. INTRODUCTION

ONE of the most important requirements imposed on new scintillators is a high energy resolution for gamma ray detection. There are two fundamental factors that determine energy resolution: Poisson statistics in the number of detected photons and the nonproportionality of the light yield of scintillators with gamma-ray energy. Nonproportionality means that the total light output of a scintillator is not precisely proportional to the energy of the absorbed gamma-ray photon. This has a deteriorating effect on energy resolution. As the light yield and the PMT performance is already close to optimal we need to reduce nonproportionality in order to improve energy resolution, and for that we wish to understand the causes of nonproportionality.

The aim of this paper is to overview, systematize, analyze and interpret the data on the photon nonproportional response (photon-nPR) of inorganic scintillation materials. The main trends and patterns of the photon-nPR typical for the different groups of scintillators, especially for oxides and halides, are highlighted. The shape of the photon-nPR curve and the degree of the photon-nPR is studied as a function of chemical composition, dopant type, refractive index and other properties of the materials. To analyze trends and patterns of nonproportionality the photon-nPRs of 22 inorganic scintillators were measured using synchrotron irradiation in the energy range from 9 to 100 keV. The data on the photon-nPRs of 11 other inorganic scintillators were obtained from literature.

## II. HISTORIC OVERVIEW OF NONPROPORTIONALITY STUDIES

The scintillation response as a function of X-ray or gamma photon energy as shown in Fig.1 will hereafter be referred to as the photon-nPR. We define the photon-nPR of a scintillator at energy $E$ as the number of photoelectrons $N_{phe}^{PMT}$ per MeV of absorbed energy observed at energy $E$ divided by the number $N_{phe}^{PMT}/MeV$ observed at $E$ = 662 keV energy. The photon-nPR is expressed as a percentage value. The energy resolution is defined as the full width of the full absorption peak in the pulse height spectrum at half the maximum intensity divided by its energy. To express the discontinuity of the photon-nPR at the binding energies of the K- or L-electrons as shown in Fig. 1, we will use the name K- or L-dip.

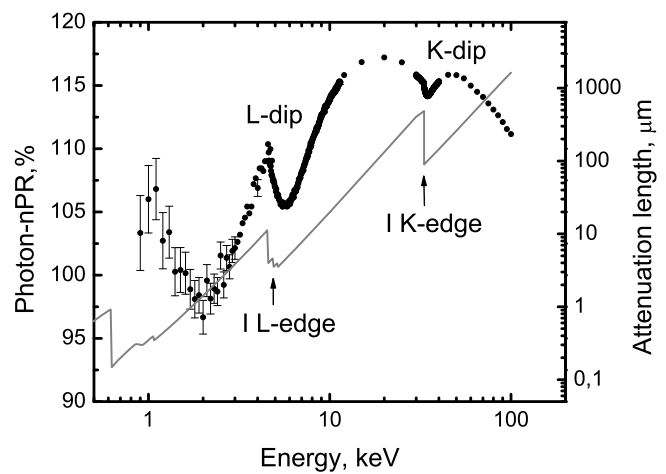

Fig. 1. Photon nonproportional response of NaI:Tl as a function of X-ray or gamma photon energy [1]. Solid line is the X-ray and gamma-ray attenuation length of NaI.

Manuscript received April 10, 2012. The research leading to these results has received funding from the Netherlands Technology Foundation (STW), Saint Gobain crystals and detectors division, Nemours, France, and the European Community's Seventh Framework Programme (FP7/2007-2013) under grant agreement n° 226716.

Ivan V. Khodyuk and Pieter Dorenbos are with the Luminescence Materials Research Group, Faculty of Applied Sciences, Delft University of Technology, Mekelweg 15, Delft, 2629JB, The Netherlands (e-mails: i.v.khodyuk@tudelft.nl, p.dorenbos@tudelft.nl).

Despite a long history of nonproportionality studies, the number of publications with systematic studies trying to identify trends is fairly small. One of the first comprehensive studies was done by Aitken et. al in 1967 [2]. They investigated the photon-nPRs of NaI:Tl, CsI:Tl, CsI:Na and $CaF_2$:Eu. The three iodides show qualitatively similar photon-nPRs down to a photon energy of about 20 keV. The shape of the calcium fluoride photon-nPR curve differs significantly. The only common feature noted was a dip in the photon-nPR near K-shell and L-shell absorption edges of iodine or calcium.

With an increasing number of available scintillators, the amount of data on nonproportionality also increased. In 1995 an overview paper by Dorenbos et. al was published [3]. It describes the photon-nPRs of "classic" (NaI:Tl, CsI:Tl, CsI:Na, $CaF_2$:Eu, $Bi_4Ge_3O_{12}$ (BGO), and $CdWO_4$ (CWO)), and "modern" ($BaF_2$, $Gd_2SiO_5$:Ce (GSO:Ce), $YAlO_3$:Ce (YAP:Ce), $Lu_2SiO_5$:Ce (LSO:Ce), $Lu_3Al_5O_{12}$:Sc (LuAG:Sc), and $K_2LaCl_5$:Ce) scintillators. The photon-nPR curves of LSO:Ce samples with different light output and energy resolution were measured. Based on these results it was shown that the photon-nPR of LSO:Ce does not depend or very weakly on the parameters that determine scintillator's quality, such as: impurities, defects, synthesis parameters, self-absorption, concentration of vacancies, presence of the afterglow, etc. A similar study to establish a relationship between the energy resolution, light output and afterglow of LSO:Ce with photon-nPR was carried out by Kapusta et. al in 2005 [4]. The authors suggested a correlation between presence of traps, photon-nPR and intrinsic energy resolution. Nevertheless, a direct relationship was not established, and most proportional response at room temperature was shown by the LSO:Ce crystal with the lowest light yield and energy resolution of about 16.7% at 662keV.

The next important step in the systematization of nonproportionality was done by Rooney, Valentine and co-workers [5-8]. Using Compton Coincidence Technique, the electron-nPR curves of the most common scintillators at that time were studied as a function of Compton electron energy. The major trend was that iodide compounds show a different shape of the electron-nPR curve than other types of scintillators, mostly oxides. High proportionality of YAP:Ce electron-nPR was also highlighted.

Belcerzyk et. al [9] studied the oxides YSO:Ce, GSO:Ce, LSO:Ce, and $Lu_{1.8}Gd_{0.2}SiO_5$:Ce (LGSO:Ce). They concluded that the shape of the photon-nPR is mainly determined by the crystal structure, but not by the type of rare earth cation. The photon-nPR of the LGSO:Ce crystal, in which 10% of the Lu atoms, were substituted by Gd atoms demonstrated a stronger nonproportionality in comparison with LSO:Ce and GSO:Ce.

In 2002, Dorenbos continued to systematize data on the photon-nPR [10] and introduced a new quantity, the so-called degree of nonproportionality. A clear correlation between the degree of nonproportionality and the deviation from the fundamental limit of energy resolution was noted.

Cutler et. al [11] studied the dependence of photon-nPR of lutetium- and yttrium-based silicates and aluminates (LSO:Ce; LSO:Ce,Ca; $Lu_2Si_2O_7$:Ce (LPS:Ce), LuAG:Pr; YSO:Ce and YSO:Ce,Ca) on crystal structure, crystal growth atmosphere, activator concentration and type of co-doping. They established that the photon-nPR of Lu- and Y-based scintillators is not significantly affected by activator concentration or substitution of crystal matrix rare earth (Lu by Y and vice versa). In addition, the authors observed that the photon-nPR depends on the composition; LSO:Ce appears more proportional than LPS:Ce, and aluminates are generally more proportional than silicates, except for LuAP:Ce. They also observed that the synthesis conditions and type of co-doping did not significantly influence the shape and degree of the photon-nPR. The usage of $Ca^{2+}$ co-doping did not have any significant effect on the photon-nPR of LSO:Ce.

In 2009, Swiderski et. al [12] explored the photon-nPR of LuAG:Pr, LuAG:Ce, LSO:Ce and $LaBr_3$:Ce scintillators versus crystal structure and composition properties and dopant type. It was shown that LuAG:Pr is more proportional than LuAG:Ce, and LuAG:Ce in turn is more proportional than LSO:Ce.

A large contribution was made by Payne et. al [13, 14]. Using SLYNCI (Scintillator Light Yield Nonproportionality Compton Instrument), the electron-nPRs of 29 scintillators were measured in the energy range from 3 to 460 keV. The data obtained by the authors were compared with the results of calculations using a model they proposed. The model is based on theories by Onsager, Birks, Bethe-Bloch, Landau and by appropriate choice of parameters it can accurately reproduce the experimental data. Seven groups of scintillator materials were distinguished: "alkali halides, simple oxides, silicates, fluorides, organics, multivalent halides and Gd-based compounds".

In addition, papers with a more theoretical approach to the problem appeared. Like by Murray et. al [15], Lempicki et. al [16], Rodnyi et. al [17, 18], Bizarri et. al [19], Keresit et. al [20], Setyawan et. al [21], Vasil'ev [22], Singh [23], Li et. al [24], Williams et. al [25].

### III. EXPERIMENTAL METHODS

To determine the photon-nPR of scintillators, scintillation pulse height spectra at many finely spaced X-ray energy values between 9 keV and 100 keV at the X-1 beamline at the Hamburger Synhrotronstrahlungslabor (HASYLAB) synchrotron radiation facility in Hamburg, Germany were carried out. The scheme of the experimental set-up can be found in [1, 26]. A monochromatic pencil X-ray beam was used as an excitation source. A tunable double Bragg reflection monochromator using a Si[511] set of silicon crystals providing an X-ray resolution of 2 eV at 9 keV rising to 20 eV at 100 keV was used to select the X-ray energies. The beam spot size was set by a pair of precision stepper-driven slits, positioned immediately in front of the sample coupled to the PMT. The intensity of the synchrotron beam was reduced in order to avoid pulse pileup. Lead shielding was used to protect the sample from receiving background irradiation which otherwise appeared as a broad background in our pulse height spectra.

To record synchrotron X-ray pulse height spectra of a scintillator, a Hamamatsu R6231-100 PMT connected to a homemade preamplifier, an Ortec 672 spectroscopic amplifier and an Amptek 8000A multichannel analyzer were used. The crystal was optically coupled to the window of the PMT with Viscasil 60 000 cSt from General Electric. The crystal was covered with several layers of ultraviolet reflecting Teflon tape (PFTE tape) forming an "umbrella" configuration [27]. Scintillation photons reflected from the photocathode are then reflected back by the umbrella thus enhancing detection efficiency. All measurements were carried out at room temperature and were repeated several times.

The number of photoelectrons $N_{phe}^{PMT}$ per MeV of absorbed energy produced in the PMT by the scintillator was determined by comparing the positions of the $^{137}$Cs 662 keV and of the $^{241}$Am 59.5 keV photopeaks in recorded pulse height spectra with the mean value of the single photoelectron pulse height spectrum. To collect as much of the emitted light as possible, the shaping time of an Ortec 672 spectroscopic amplifier was set at 10 µs.

TABLE I
PROPERTIES OF THE SCINTILLATION MATERIALS

| Scintillator | R,% at 662 keV | Light yield, photons/keV at 662 keV | Refractive index, n | Photon-nPR,% at 10keV | Degree of photon-nPR, $\sigma_{photon\text{-}nPR}$,% | Reference |
|---|---|---|---|---|---|---|
| YAP:Ce | 4.4 | 16 | 1.95 | 95.2 | 0.13 | [3, 28-30] |
| Cs$_2$LiYCl$_6$:Ce | 3.9-5.1 | 21 | | 102.2 | 0.38 | [31-33] |
| K$_2$LaCl$_5$:Ce | 5.1 | 28 | | ~95 | 0.52 | [3, 34] |
| YSO:Pr | 8 | 6 | 1.8 | 85.3 | 0.72 | [28] |
| LuAP:Ce | 7.1-9.5 | 12 | 1.95 | ~87 | 0.94 | [35-37] |
| LaBr$_3$:Ce | 2.7-3.1 | 75 | 2.05 | 87.2 | 1.10 | [12, 38-40] |
| ZnO (ceramics) | 11.8 | 25 | 2.03 | 82.7 | 1.33 | [41-43] |
| SrI$_2$:Eu | 2.8-3.7 | 85-115 | 2.05 | 96.1 | 1.35 | [14, 44, 45] |
| LuAG:Sc | 6.5 | 23 | 1.84 | ~90 | 1.36 | [3] |
| SrI$_2$ | 6.7 | 38 | 2.05 | 87 | 1.61 | [44] |
| LuAG:Pr | 4.6 | 16 | 1.84 | 86 | 1.98 | [12, 46] |
| LuI$_3$:Ce | 3.3 | 98 | | 85 | 2.12 | [47] |
| LaCl$_3$:Ce | 3.1 | 49 | 1.9 | 83.7 | 2.16 | [38, 39, 48] |
| CeBr$_3$ | 4-4.4 | 68 | | ~80 | 2.42 | [49] |
| LuYAP:Ce | 8-10 | 15 | 1.94 | ~75 | 2.77 | [36, 50, 51] |
| BaF$_2$ | 7-8 | 11 | 1.51 | 80 | 3.54 | [3, 52, 53] |
| YAG:Ce | 3.5-15 | 17 | 1.82 | ~82 | 3.64 | [54, 55] |
| LuAG:Ce | 11 | 12.5 | 1.84 | ~70 | 4.04 | [12, 54] |
| CaF$_2$:Eu | 5.7 | 24 | 1.44 | ~72 | 4.45 | [2] |
| GSO:Ce | 9.2-9.5 | 12 | 1.85 | 68.8 | 5.03 | [9, 46] |
| YAP | 13.3 | 1.5 | 1.95 | ~70 | 5.70 | [56] |
| YSO:Ce | 9.4-11.1 | 24 | 1.792 | ~65 | 6.71 | [9, 11] |
| CsI:Tl | 5.8-6.2 | 57 | 1.78 | 112 | 6.78 | [2, 5] |
| NaI:Tl | 5.6-7.8 | 45 | 1.85 | 114 | 6.83 | [1, 2, 57] |
| YPO$_4$:Ce | | | 1.65 | 58.3 | 7.06 | [28] |
| CWO | 6.6 | 15 | 2.25 | 47.3 | 7.11 | [3, 58, 59] |
| BGO | 7.8-10 | 8 | 2.15 | ~70 | 7.15 | [3, 60, 61] |
| LSO:Ce | 7.9-11.9 | 29 | 1.82 | 56.7 | 8.40 | [3, 9, 12, 46, 62] |
| CsI:Na | 6.5 | 49 | 1.84 | ~130 | 8.46 | [2] |
| LPS:Ce | 10 | 26 | 1.74 | ~42 | 8.94 | [46, 63] |
| LYSO:Ce,Ca | 8.7 | 38 | 1.81 | ~45 | 11.24 | [64] |
| LGSO:Ce | 12.4 | 20 | | 45 | 12.56 | [9] |
| NaI | 16.2-18.1 | 1-3 | 1.80 | ~170 | 18.14 | [65] |

## IV. RESULTS

In this section data on the photon-nPR of 33 inorganic scintillation materials are presented. The photon-nPRs of 22 inorganic scintillators were measured using synchrotron irradiation in the energy range from 9 to 100 keV at HASYLAB. Data on the photon-nPR of 11 other inorganic scintillators were obtained from published works.

All investigated scintillators and their properties are listed in Table I. In addition to properties, such as energy resolution, light yield and refractive index, Table I compiles the photon-nPR at 10 keV and the degree of photon-nPR ($\sigma_{photon\text{-}nPR}$). If the exact value of the photon-nPR at 10 keV has not been determined in the measurement, we used an approximate value obtained by extrapolating the response from a fit at higher excitation energies. The degree of photon-nPR was determined following the ideas in [10, 21]. Its numerical value was calculated using

$$\sigma_{photon-nPR} = \frac{1}{(E_{max} - E_{min})} \times \\ \times \int_{E_{min}}^{E_{max}} |f_{photon-nPR}(E_{max}) - f_{photon-nPR}(E)| dE \quad (1)$$

where $E_{max}$ = 662 keV, $E_{min}$ = 10 keV, $f_{photon-nPR}(E_{max})$ = 100%, and $f_{photon-nPR}(E)$ is the value of the photon-nPR at the excitation energy E. For a perfectly proportional scintillator the value of the degree of photon-nPR is zero. As a result, scintillators with a lower value of the $\sigma_{photon-nPR}$ are considered to be more proportional.

### A. Oxide scintillators

A large number of currently known and applied scintillators are oxides. Oxides are generally nonhygroscopic dense scintillators with high gamma stopping power. According to the type of rare earth (RE) cation and chemical composition, oxides will be divided into the groups shown in Table 2.

TABLE II
CLASSIFICATION OF OXIDE SCINTILLATORS

| Chemical composition | | RE-cation | | | |
|---|---|---|---|---|---|
| | | $Y^{3+}$ | $Gd^{3+}$ | $Lu^{3+}$ | Mixed Y-Gd-Lu |
| Aluminum perovskites | $REAlO_3$ | YAP:Ce<br>YAP:Pr<br>YAP | | LuAP:Ce | LuYAP:Ce |
| Aluminum garnets | $RE_3Al_5O_{12}$ | YAG:Ce | | LuAG:Ce<br>LuAG:Pr<br>LuAG:Sc | |
| Oxyorthosilicates | $RE_2SiO_5$ | YSO:Ce<br>YSO:Pr | GSO:Ce | LSO:Ce | LYSO:Ce,Ca<br>LGSO:Ce |
| Pyrosilicates | $RE_2Si_2O_7$ | | | LPS:Ce | |
| Phosphates | $REPO_4$ | $YPO_4$:Ce | | | |

*Influence of chemical composition*

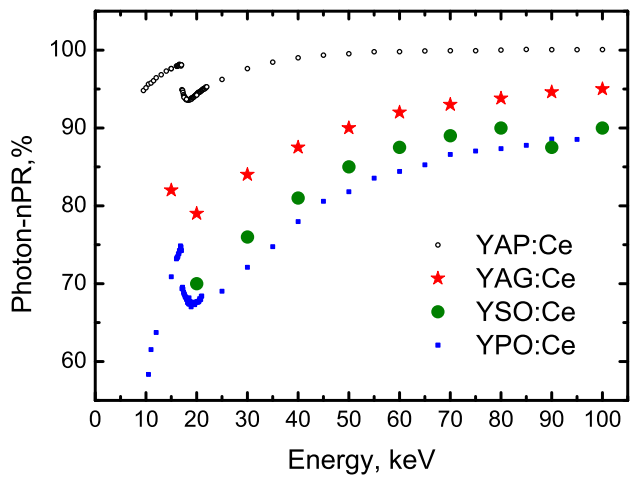

Fig. 2. Photon nonproportional response of Ce-doped Y-based scintillators as a function of X-ray or gamma photon energy. Black open dots – YAP:Ce; red stars – YAG:Ce [54]; olive circles – YSO:Ce [11]; and blue squares – $YPO_4$:Ce.

The effect of the chemical composition on the shape and degree of the photon-nPR was noted by many authors. For the majority of cerium doped oxides regardless of the RE-cation in the composition, a trend in the photon-nPR can be observed. In Fig. 2 and 3 the photon-nPR as a function of energy of the incident X-ray or gamma photon is shown. All data presented are normalized at 662 keV. The yttrium and lutetium based compaunds show similar shape of the photon-nPR curves. The most proportional response is shown by aluminum perovskites with composition $REAlO_3$:Ce, next are aluminum garnets $RE_3Al_5O_{12}$:Ce, oxyorthosilicates $RE_2SiO_5$:Ce, phosphates $REPO_4$:Ce and pyrosilicates $RE_2Si_2O_7$:Ce, where RE = Lu or Y. In contradiction to the statement in [11] that LuAP:Ce has a high degree of photon-nPR, Fig. 3 shows that it has the same relatively low degree of photon-nPR as YAP:Ce. The degree of photon-nPR increases from 0.13% for YAP:Ce up to 7.06% for $YPO_4$:Ce and from 0.94% for LuAP:Ce up to 8.94% for LPS:Ce.

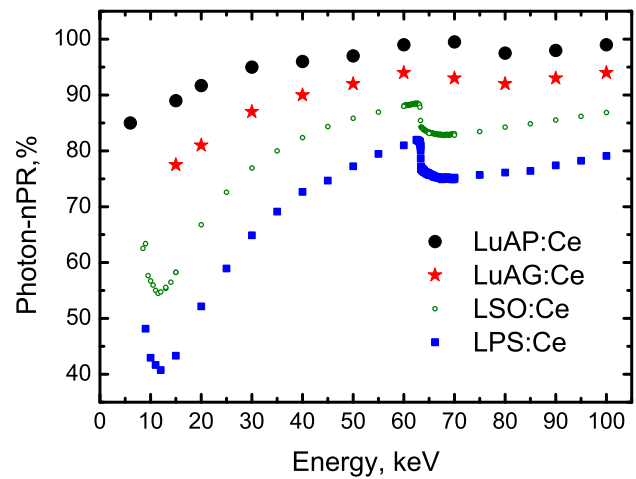

Fig. 3. Photon nonproportional response of Ce-doped Lu-based scintillators as a function of X-ray or gamma photon energy. Black circles – LuAP:Ce [35]; red stars – LuAG:Ce [12]; olive open dots – LSO:Ce; and blue squares – LPS:Ce.

Despite the strong dependence of the degree of photon-nPR on the chemical composition, the shape of the photon-nPR curves appears similar for all scintillators in Figs. 2 and 3. It is always characterized by a lowest value of the photon-nPR at the lowest energies, a dip in the photon-nPR curve near K-shell and L-shell absorption edges of yttrium or lutetium, and a smooth monotonic increase of the photon-nPR curve with increasing photon energy up to 662 keV.

*Influence of dopant*

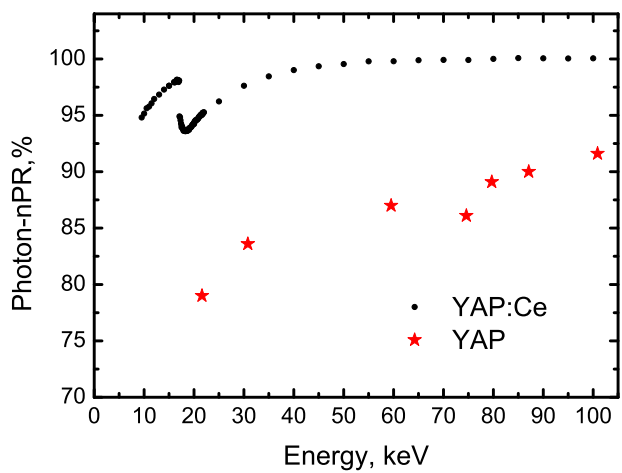

Fig. 4. Photon nonproportional response of YAP scintillators as a function of X-ray or gamma photon energy. YAP:Ce – black dots; pure YAP – red stars [56].

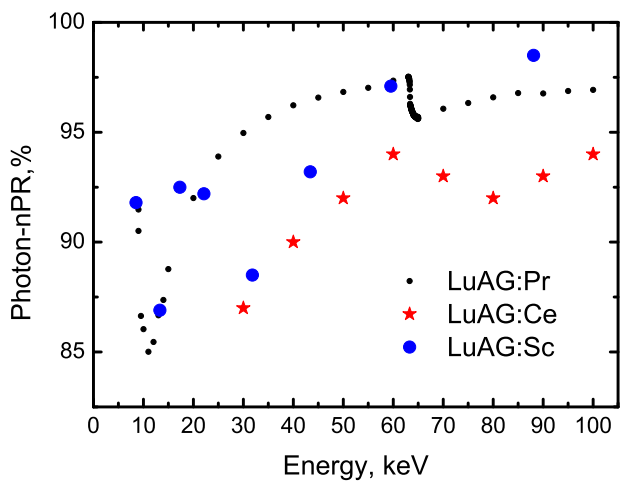

Fig. 5. Photon nonproportional response of LuAG scintillators as a function of X-ray or gamma photon energy. LuAG:Pr – black dots [46]; LuAG:Sc – blue circles [3]; LuAG:Ce – red stars [12].

The influence of dopant on nonproportionality and energy resolution was discussed in [12, 14, 18]. A clear correlation between the degree of photon-nPR and the concentration or type of the dopant was not established yet. Figures 4, 5 and 6 show the photon-nPR as a function of energy for YAP, LuAG and YSO pure materials or activated with different dopants. The most pronounced observation is the improvement of nonproportionality with $Pr^{3+}$ doping in comparison with $Ce^{3+}$ doping for LuAG and YSO. LuAG doped with praseodymium shows lower $\sigma_{photon-nPR}$ = 1.98% in comparison with the cerium doped material $\sigma_{photon-nPR}$ = 4.04%. The same was observed for YSO scintillators: 0.72% and 6.71% for praseodymium and cerium doped, respectively. A relatively good proportionality of YSO:Pr scintillator leads to the fact that even with low light yield 6000 photons/MeV this material shows energy resolution of about 8% at 662 keV.

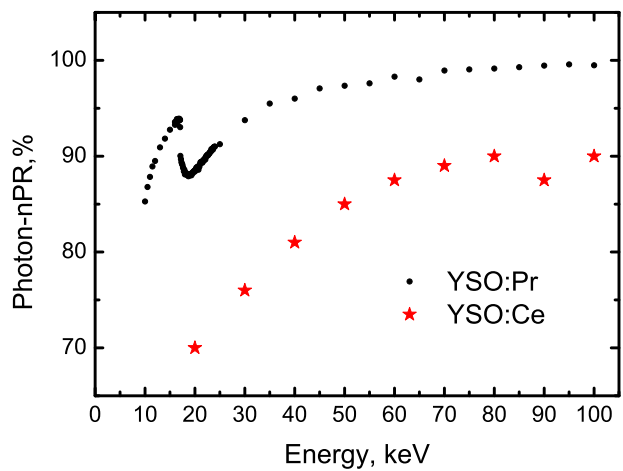

Fig. 6. Photon nonproportional response of YSO scintillators as a function of X-ray or gamma photon energy. YSO:Pr – black dots; YSO:Ce – red stars [11].

Together with the doped scintillator YAP:Ce, Fig. 4 shows the photon-nPR curve for undoped YAP [56]. Undoped YAP shows much larger $\sigma_{photon-nPR}$ = 5.70% then that of Ce-doped material $\sigma_{photon-nPR}$ = 0.13%. The light yield of the pure material is less than 10% of the light output of YAP:Ce, which leads to a substantial contribution made to the overall energy resolution 13.3% at 662 keV by photo-detector statistics [3].

Figure 5 shows the photon-nPR as a function of X-ray or gamma energy for LuAG crystals doped not only with $Ce^{3+}$ and $Pr^{3+}$, but also with $Sc^{3+}$. In the case of Pr-doped scintillator the photon-nPR is more proportional in comparison with LuAG:Ce. The values of the LuAG:Sc photon-nPR with the exception of two values at energies of 30 and 45 keV almost coincide with the numbers of the LuAG:Pr photon-nPR. The most proportional LuAG:Sc with $\sigma_{photon-nPR}$ = 1.36% and LuAG:Pr with $\sigma_{photon-nPR}$ = 1.98% have energy resolutions of 6.5% and 4.6% at 662 keV, respectivey, whereas less proportional LuAG:Ce with $\sigma_{photon-nPR}$ = 4.04% shows energy resolution of 11% at 662 keV.

*RE-cation substitution effect*

To improve scintillation properties like afterglow, decay time and light yield, and to improve stability of the crystal during the growth process, fraction of the RE-cations in a material composition can be replaced by RE-cations of lower atomic number. The most widely known and used material of this kind is LYSO:Ce in which about 10% of the lutetium in LSO:Ce is replaced by yttrium. Sometimes in order to improve the scintillation properties of LYSO:Ce co-doping with calcium or magnesium can be used [64]. Another well-known example of a scintillator with mixed RE-cation is LGSO:Ce [9].

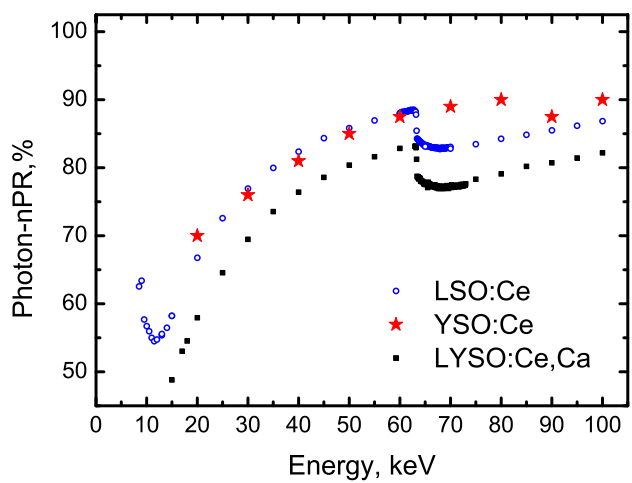

Fig. 7. Photon nonproportional response of Ce-doped oxyorthosilicates as a function of X-ray or gamma photon energy. LSO:Ce – blue open dots; YSO:Ce – red stars [11]; LYSO:Ce,Ca – black squares.

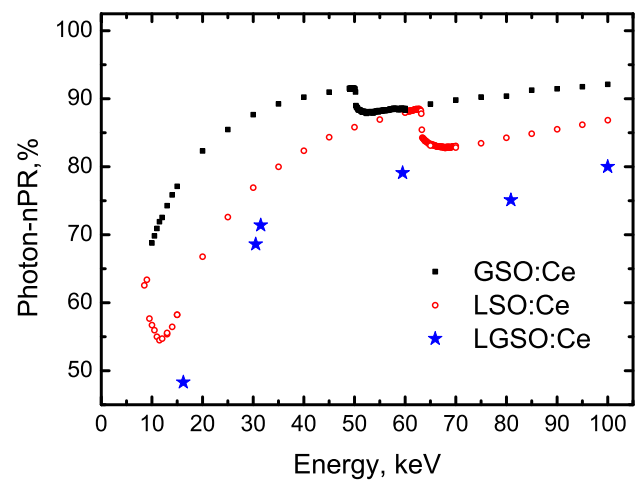

Fig. 8. Photon nonproportional response of Ce-doped oxyorthosilicates as a function of X-ray or gamma photon energy. GSO:Ce – black squares; LSO:Ce – red open dots; LGSO:Ce – blue stars [9].

Figure 7 compares the LYSO:Ce,Ca photon-nPR with that of LSO:Ce and YSO:Ce. A similar comparison of the LGSO:Ce photon-nPR with that of LSO:Ce and GSO:Ce is shown in Fig. 8. The mixed RE-cation compounds, show a high degree of photon-nPR: 11.24% and 12.56% for LYSO:Ce,Ca and LGSO:Ce, respectively. An effect similar to the photon-nPR deterioration for the mixed-cation compounds can be observed for aluminum perovskites, as shown in Fig. 9. In this figure, the photon-nPR of LuYAP:Ce is compared with the photon-nPRs of YAP:Ce and LuAP:Ce scintillators. YAP:Ce and LuAP:Ce have low $\sigma_{photon-nPR}$ = 0.13% and 0.94%, compared with other oxides. Whereas the degree of LuYAP:Ce photon-nPR is 2.77%. Similar to the case of oxyorthosilicates, shown in Figs. 7 and 8, and similar to the case of aluminum perovskites, shown in Fig. 9, the complete replacement of yttrium or gadolinium for lutetium did not significantly change the degree of photon-nPR. Only a slight improvement of $\sigma_{photon-nPR}$ can be noticed for GSO:Ce (5.03%) and YAP:Ce (0.13%).

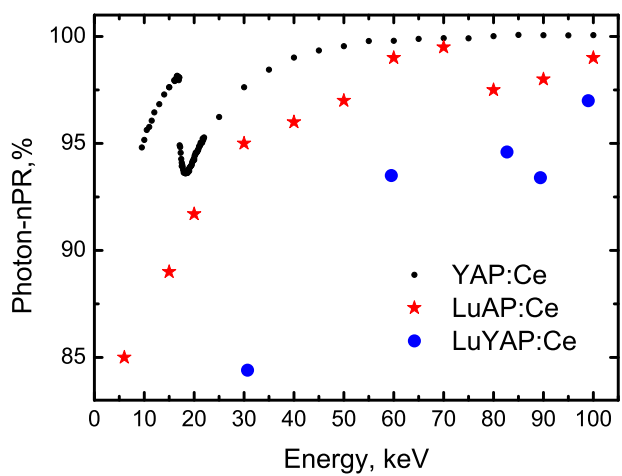

Fig. 9. Photon nonproportional response of Ce-doped aluminum perovskites scintillators as a function of X-ray or gamma photon energy. YAP:Ce – black dots; LuAP:Ce – red stars [35]; LuYAP:Ce – blue circles [50].

*Other oxides*

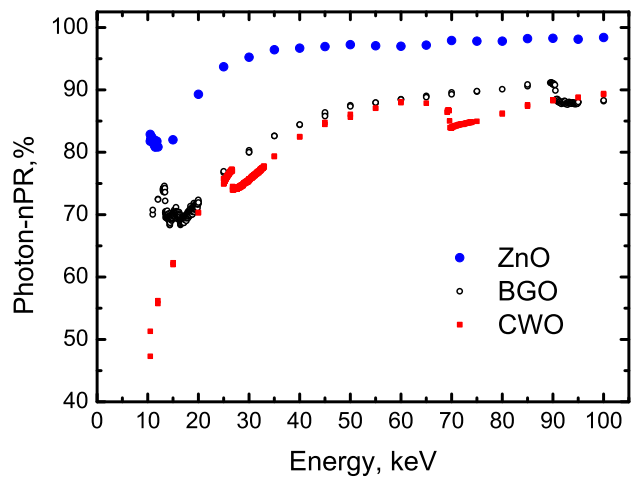

Fig. 10. Photon nonproportional response of oxide scintillators as a function of X-ray or gamma photon energy. ZnO – blue circles [66]; BGO – black open dots; CWO – red squares.

Fig. 10 shows the photon-nPR curves for ceramic ZnO, BGO and CWO. ZnO shows the smallest degree of photon-nPR $\sigma_{photon-nPR}$ = 1.33. The ceramic sample was 1.5 mm thick with a transmittance of about 50% at the maximum of the luminescence wavelength [41]. Synchrotron X-ray excitation was on one side of the sample and the registration of the luminescence on the other [1]. Part of the light, especially at low X-ray energy, can be self-absorbed, which may lead to an increased downturn in the photon-nPR at low excitation energies than in fully transparent ZnO. As shown in Fig. 10 the photon-nPRs of BGO and CWO show similar shape and degree of photon-nPR above 7%. With a light output equal to 8000 photons/MeV for BGO and 15000 photons/MeV for CWO and the observed photon-nPRs, the energy resolution of these materials is about 8% and 7%, respectively.

## B. Halide scintillators

Halide scintillators play an important role among the presently known scintillation materials. Doped with rare-earth ions $Ce^{3+}$ and $Eu^{2+}$ they are especially important. In this section the effect of anion and cation on the shape and degree of photon-nPR of fluorides, chlorides, bromides and iodides is considered. Also the photon-nPR of pure iodides, iodides doped with $Tl^+$, $Na^{2+}$ or $Eu^{2+}$ is compared.

### *Anion effect (F – Cl – Br – I)*

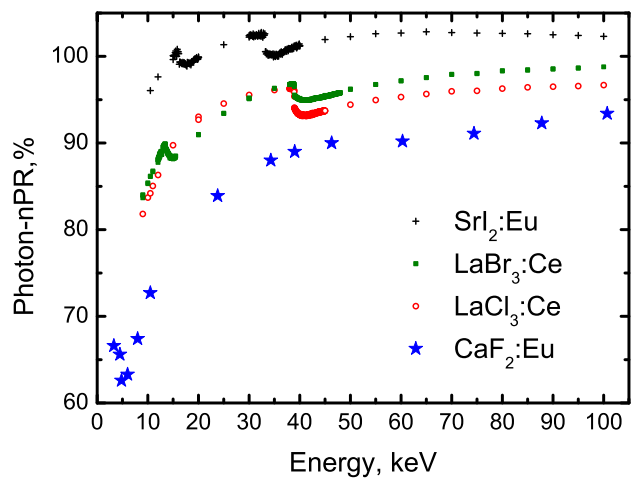

Fig. 11. Photon nonproportional response of halide scintillators as a function of X-ray or gamma photon energy. $SrI_2$:Eu – black crosses [44]; $LaBr_3$:Ce – olive squares [39]; $LaCl_3$:Ce – red open dots [39]; $CaF_2$:Eu – blue stars [2].

Figure 11 shows the photon-nPRs of $SrI_2$:$Eu^{2+}$, $LaBr_3$:$Ce^{3+}$, $LaCl_3$:$Ce^{3+}$, and $CaF_2$:$Eu^{2+}$. The concentration of $Eu^{2+}$ in $SrI_2$:Eu was 5% [44], while for $LaBr_3$:Ce and $LaCl_3$:Ce the standard commercially available scintillators BrilLanCe 380 and BrilLanCe 350 containing 5% and 10% of $Ce^{3+}$ were used. The $CaF_2$:Eu data were taken from the paper by Aitken et. al [2], where the concentration of $Eu^{2+}$ was not specified.

Table I shows that $LaBr_3$:Ce has the lowest degree of photon-nPR 1.10%. The value of the photon-nPR at 10 keV increases from fluoride to chloride to bromide to iodide. The photon-nPRs of all investigated scintillators were normalized to the corresponding value at 662 keV. If normalization is at a different energy, e.g. at 100keV, $SrI_2$:Eu will show the lowest degree of photon-nPR. In the energy range 10 – 100 keV, the $SrI_2$:Eu photon-nPR is always within 8% from 100%, that is twice smaller than for $LaBr_3$:Ce. Among the scintillators presented in Fig. 11, $SrI_2$:Eu is the only material with the photon-nPR values exceeding 100% in the energy range 10-100 keV.

### *Fluorides*

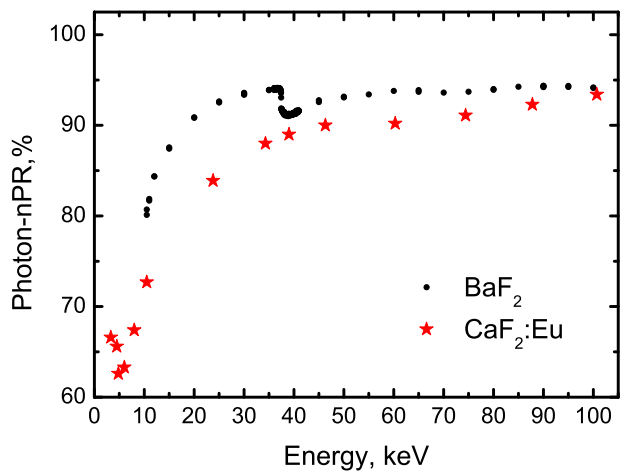

Fig. 12. Photon nonproportional response of fluoride scintillators as a function of X-ray or gamma photon energy. $BaF_2$ – black dots; $CaF_2$:Eu – red stars [2].

Comparing the photon-nPR, that we measured for pure $BaF_2$ and the photon-nPR of $CaF_2$:Eu, taken from the reference [2], it can be seen that $BaF_2$ is a more proportional scintillator. $BaF_2$ shows two types of luminescence; the core to valence luminescence (CVL) showing peaks at 195 and 220 nm [67-69] and excitonic luminescence with a broad band peaking around 310 nm. The quantum efficiency of the Hamamatsu R6231-100 PMT that was used is low below 300 nm [70]. Thus Fig. 12 shows the photon-nPR mainly for the excitonic luminescence of $BaF_2$.

### *Chlorides*

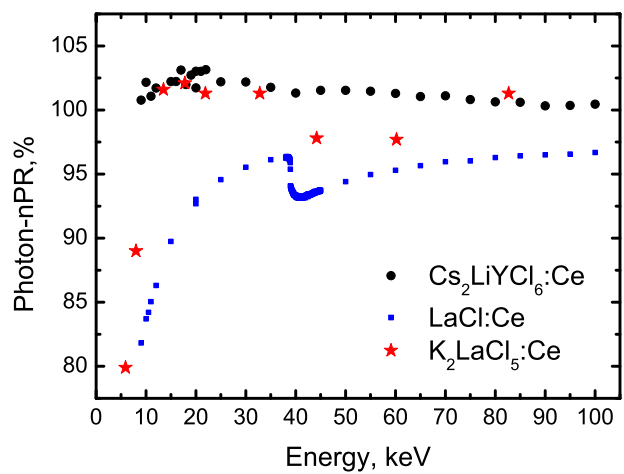

Fig. 13. Photon nonproportional response of chloride scintillators as a function of X-ray or gamma photon energy. $Cs_2LiYCl_6$:Ce – black dots; $K_2LaCl_5$:Ce – red stars [3]; $LaCl_3$:Ce – blue squares [39].

Fig. 13 shows shows photon-nPRs for chloride scintillators. The data on $K_2LaCl_5$:Ce were taken from [3]. CLYC:Ce was studied at the end of 90's [71, 72] and in the early 2000's [31, 73]. It is a very proportional scintillator with small degree of photon-nPR 0.38% making it one of the most proportional scintillators. Only YAP:Ce with $\sigma_{photon\text{-}nPR}$ = 0.13% appears better. $K_2LaCl_5$:Ce also demonstrates small photon-nPR except at low excitation energies of about 6 and 10 keV. However, these data points contain large errors. As for the photon-nPR of LaCl$_3$:Ce, it has the highest $\sigma_{photon-nPR}$ = 2.16% among the three investigated chlorides.

Nonproportionality is just one of the factors affecting the total energy resolution of a scintillator. So the even less proportional material LaCl$_3$:Ce with higher light yield, shows an energy resolution of 3.1% at 662 keV, while the more proportional materials CLYC:Ce and K$_2$LaCl$_5$:Ce both show only 5.1% at 662 keV. From the standpoint of the highest possible energy resolution it should always be the optimal combination of low nonproportionality and high light output.

*Bromides*

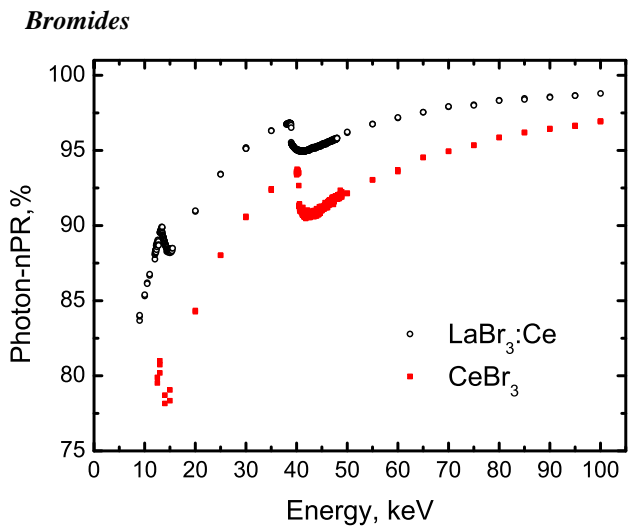

Fig. 14. Photon nonproportional response of bromide scintillators as a function of X-ray or gamma photon energy. LaBr$_3$:Ce – black open dots [39]; CeBr$_3$ – red squares.

Figure 14 shows the photon-nPR of LaBr$_3$:Ce and CeBr$_3$ scintillators. CeBr$_3$ shows high light yield, but the 4% energy resolution at 662 keV is worse than that of LaBr$_3$:Ce. As can be seen in Fig. 14, poorer energy resolution of CeBr$_3$ can be related to the higher degree of photon-nPR $\sigma_{photon-nPR}$ = 2.42% as compared with $\sigma_{photon-nPR}$ = 1.10% of LaBr$_3$:Ce.

*Iodides*

Figure 15 shows photon-nPRs of CsI:Tl, NaI:Tl, SrI$_2$:Eu and LuI$_3$:Ce. The data on the CsI:Na photon-nPR were taken from [2]. The photon-nPR curves in Fig. 15 vary a lot. LuI$_3$:Ce has a shape more similar to LaBr$_3$:Ce than to NaI:Tl. SrI$_2$:Eu shows a low degree of photon-nPR, the 1.35%. Its photon-nPR curve runs between that of LuI$_3$:Ce and NaI:Tl. CsI:Tl, NaI:Tl and CsI:Na, show the typical shape of the photon-nPR for iodides [2] with a maximum near 10 - 20 keV.

Figure 16 compares the photon-nPRs for pure SrI$_2$ and NaI, with that of SrI$_2$:Eu and NaI:Tl at room temperature. For SrI$_2$ the change of the photon-nPR with the introduction of Eu$^{2+}$ is relatively small. As seen in Table 1, doping by Eu$^{2+}$ substantially increases the light output of SrI$_2$, but the degree of photon-nPR is only slightly reduced from 1.61% to 1.34%. For NaI the photon-nPR improves considerably by doping with Tl$^+$. Figure 16 shows that the value of the photon-nPR of pure NaI reaches 170% at 17 keV X-ray energy. In fact, this means that the energy conversion efficiency of 17 keV X-ray or gamma radiation into the optical photons is 1.7 times higher than that of 662 keV. This leads to the largest degree of photon-nPR, 18.14%, among all the scintillators considered in this paper. The large degree of photon-nPR of pure NaI, combined with low light yield leads to an energy resolution of 16.2 – 18.1% at 662 keV [65].

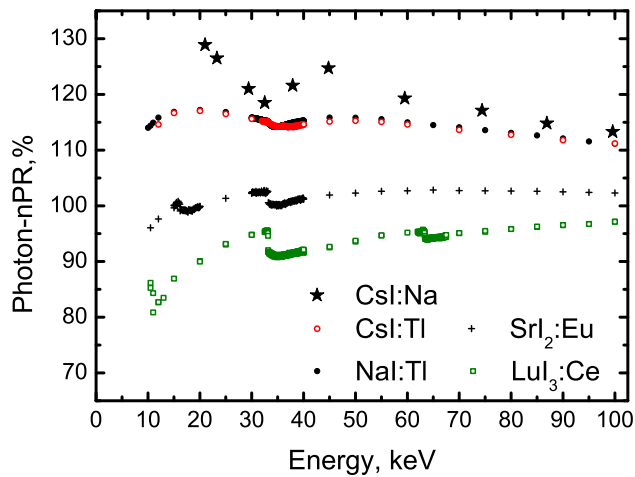

Fig. 15. Photon nonproportional response of iodide scintillators as a function of X-ray or gamma photon energy. CsI:Na – black stars [2]; NaI:Tl – black dots [1]; CsI:Tl – red open dots; SrI$_2$:Eu – black crosses [44]; LuI$_3$:Ce – olive open squares [30, 39].

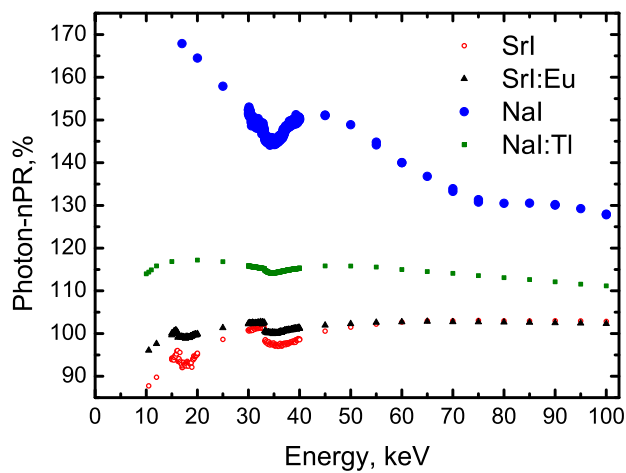

Fig. 16. Photon nonproportional response of iodde scintillators as a function of X-ray or gamma photon energy. NaI – blue ciricles; NaI:Tl – olive squares [1]; SrI$_2$:Eu – black triangles [44]; SrI$_2$ – red open dots [44].

V. DISCUSSION

Several models were proposed to explain the origin of nonproportionality and the shape of the photon- and electron-nPR curves [14, 19, 20, 23, 25]. In all those models, the cause of nonproportionality is a high concentration of charge carriers resulting from the interaction of ionizing radiation

> REPLACE THIS LINE WITH YOUR PAPER IDENTIFICATION NUMBER (DOUBLE-CLICK HERE TO EDIT) < 9

with a scintillator material. According to the Bethe equation, with decrease in the energy of the primary electron, the ionization density ($dE/dx$) along the track grows [74]. This leads to larger radiationless electron hole recombination rate which forms the basis of increasing nonproportionality with smaller X-ray or gamma-ray energy. This increase can be clearly observed for the scintillators investigated in this study by the downward curvature of the photon-nPR curves when going to smaller energy. Exceptions are CLYC:Ce shown in Fig. 13, CsI:Na shown in Fig. 15, and NaI shown in Fig. 16. Probably their nonproportionality curve does not differ fundamentally from that of the others; the only difference is that the dropping of the photon-nPR starts at X-ray energy below 10 keV that is out of the scope of our experiments. The models presented use slightly different arguments to explain the degree of photon- or electron-nPR. A comprehensive overview of the currently available models and ideas was presented by Moses et. al [75]. The basis of all those models is the competition between two opposing processes: 1) quenching due to radiationless electron hole recombination inside the volume of high ionization density along the track, and 2) diffusion of the charge carriers from the point of creation towards a volume of lower ionization density. The faster the charge carriers escape the volume of high ionization density in which quenching occurs, the higher the probability of converting the energy of the carriers into optical photons. An important factor determining the rate at which carriers leave this volume is the carrier diffusion coefficient [24, 25, 76]. A high diffusion coefficient contributes to a more rapid transport of electrons, holes and excitons to regions further from the track where the radiationless recombination rate does not depend on ionization density.

According to the diffusion equation [77]:

$$\frac{\partial n(r,t)}{\partial t} = \nabla\left[D(n)\nabla n(r,t)\right], \quad (2)$$

where $n$ is the concentration of charge carriers, $r$ is the radial coordinate perpendicular to the ionization track, $t$ is the time and $D$ is the diffusion coefficient. Assuming that in the high ionization density volume the diffusion coefficient is independent of concentration: $D(n) = const$ [78], and using the Einstein relation $D = \mu \cdot \frac{kT}{e}$, one obtains

$$\left.\frac{\partial n(r,t)}{\partial t}\right|_{diffusion} = \mu \cdot \frac{kT}{e} \cdot \nabla^2 n(r,t), \quad (3)$$

where $\mu$ is the mobility of carriers, $k$ is the Boltzmann constant, $T$ is the effective temperature of the carriers and $e$ is the elementary charge.

Based on Eq. (3), the transport of the charge carriers is faster when mobility and temperature increases. Hence, the photon-nPR is expected to depend on carrier mobility and temperature. Such dependence on temperature was indeed observed for LaBr$_3$:Ce and SrI$_2$:Eu [44, 79, 80].

According to theory [78, 81], the mobility for thermalized carriers in wide band gap semiconductors is determined by lattice scattering and impurity scattering. The mobility determined by the lattice scattering mechanism $\mu_L$ is given by:

$$\mu_L = \frac{e}{2\alpha\omega_0 m^*}\left(\exp\left(\frac{\hbar\omega_0}{kT}\right) - 1\right), \quad (4)$$

$$\alpha = \left(\frac{1}{\varepsilon_\infty} - \frac{1}{\varepsilon}\right)\sqrt{\frac{m^* E_H}{m_e \hbar\omega_0}}$$

where $\alpha$ is the polaron coupling constant, $\omega_0$ is the longitudinal optical phonon frequency, $m^*$ is the effective mass of the carriers – holes or electrons, $\hbar$ is the reduced Planck constant, $m_e$ is the rest mass of the electron, $E_H$ is the first ionization energy of the hydrogen atom (13.595 eV), $\varepsilon_\infty$ and $\varepsilon$ are the high frequency and the static dielectric permittivity constants [78]. In turn, the mobility determined by the impurity scattering mechanism $\mu_I$ is given by:

$$\mu_I = \frac{2^{7/2}(\varepsilon\varepsilon_0)^2 (kT)^{3/2}}{\pi^{3/2} z^2 e^3 (m^*)^{1/2} N_i} \cdot F(3kT), \quad (5)$$

where $z$ is the effective charge of the impurity with concentration $N_i$ and $\varepsilon_0$ is the vacuum permittivity and $F(3kT)$ is the averaged Coulomb screening factor [81].

Equations (4) and (5) are valid for thermalized carriers in wide band gap semiconductors. In the case of charge carrier transfer from the primary track to the luminescence centers in scintillators, the picture can be different. For example it is not clear if we can speak about thermalized carriers or if extra kinetic energy should be taken into account. Anyway, in both the lattice scattering mechanism Eq. (4) and the impurity scattering mechanism Eq. (5) the carrier mobility increases with increasing value of the high frequency or static dielectric permittivities of the material. Unfortunately, the dielectric permittivities are not known for all scintillators discussed in this paper. However the refractive index is well known for the scintillators. Refractive indexes are listed in Table I. The high frequency refractive index is related to the dielectric permittivity as

$$n = \sqrt{\varepsilon_r \mu_r}, \quad (6)$$

where the relative permittivity is given by $\varepsilon_r = \frac{\varepsilon(\omega)}{\varepsilon_0}$ and $\mu_r$ is the relative permeability. For the majority of scintillators $\mu_r \approx 1$; and, consequently, $n^2 \sim \varepsilon(\omega)$. For higher values of the refractive index there is a higher dielectric permittivity and, consequently, a higher carrier mobility, faster diffusion of carriers from the high ionization density volume of the track, lower radiationless electron hole recombination and as a result a better proportionality. Figure 17 shows the degree of photon-nPR versus the refractive index of the Ce-doped RE oxides from Table II and Figs. 1, 2. With increasing refractive index the degree of photon-nPR is decreasing. A similar dependence is observed for halide scintillators, shown in Fig. 18.

Figure 17 shows that dependence of the photon-nPR degree on the refractive index is less steep for the Y-based complex oxide scintillators compared to the Lu-based scintillators. This can be interpreted as a lower ionization density due to longer ionization track in Y-based scintillators. The average length of the ionization tracks at equal energy of the incident X-ray or gamma photon is inverse proportional to the density of the compound and its effective atomic number. The longer the ionization track the lower the high ionization density volume. This leads to a decrease in the radiationless electron hole recombination and reduced degree of photon-nPR.

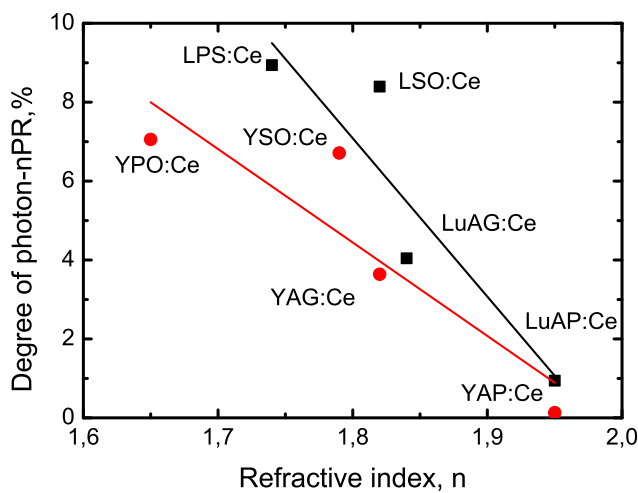

Fig. 17. Degree of photon-nPR as a function of refractive index of the oxides. Solid lines – linear approximation.

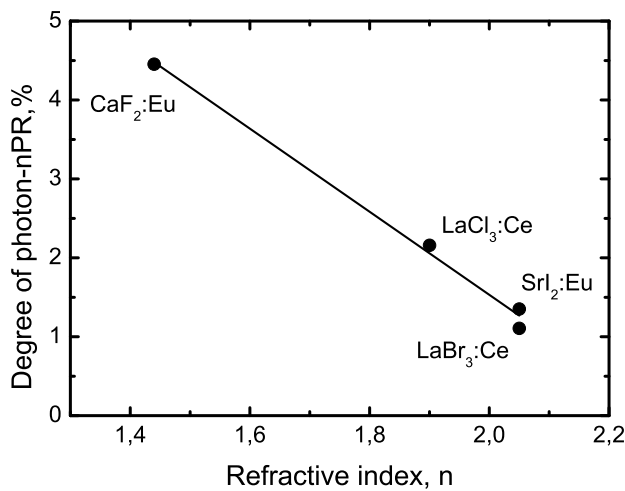

Fig. 18. Degree of photon-nPR as a function of refractive index of the halides. Solid line – linear approximation.

The high frequency dielectric permittivity and hence the refractive index are determined by how strongly valence electrons are bonded to atoms in a scintillator material. The lower the bonding energy of the electrons, the higher the dielectric permittivity and the refractive index and thus the lower the degree of photon-nPR. This rule is not absolute, and as can be seen from Table 1 and Fig. 10, some oxide scintillators with high refractive index show a relatively high degree of photon-nPR, e.g. BGO $\sigma_{photon-nPR}$ = 7.15% and CWO $\sigma_{photon-nPR}$ = 7.15%. However, when considering the general patterns of nonproportionality attention should be paid to the refractive index, dielectric constant, carriers mobility and bonding energy of the valence electrons.

The high frequency dielectric constant and the refractive index depend on the electrons in the compound. The larger the concentration of electrons and the weaker they are bonded the larger the refractive index tends to be. The weaker bonded electrons are the valence band electrons and they contribute most to the refractive index. Within the halides the valence band electrons bonding decreases in the sequence fluorides-chlorides-bromides-iodides, and within the oxides bonding depends on the cations like P, S, Si, Al that bind the oxygen ligands. Since a relationship exists between valence band electron binding and bandgap one, the refractive index tends to increase with smaller bandgap.

According to [21] the nonproportionality is influenced not only by the mobility of electrons but also by the mobility of holes. Hole mobility is related to the width of the valence band that tends to increase from fluorides to chlorides to bromides and to iodides [82]. At the higher width of the valence band the probability for hole to be trapped is lower, thus the probability of radiative recombination is higher. In the sequence fluorides-chlorides-bromides-iodides the width of the band gap on average decreases, while the width of the valence band increases, thereby resulting in higher hole mobility and better proportionality. So in Fig. 18 decrease of $\sigma_{photon-nPR}$ is explained not only by increasing refractive index, but also by increasing width of the valence band.

VI. CONCLUSION

Complex oxides show a decrease in the degree of photon-nPR in the sequence $RE_2Si_2O_7$:Ce – $REPO_4$:Ce – $RE_2SiO_5$:Ce – $RE_3Al_5O_{12}$:Ce – $REAlO_3$:Ce, where RE is Y or Lu. The decrease of the photon-nPR degree is correlated with an increase in the refractive index of the compounds. Refractive index is related to dielectric constant, carrier mobility and bonding energy of electrons. A decrease in the degree of photon-nPR similar to the complex oxides was observed for the halide scintillators with the anion replacement: fluoride to chloride to bromide and to iodide. In some materials the dopant has influence on the shape and degree of the photon-nPR. So for LuAG and YSO hosts a decrease in the degree of photon-nPR was observed after doping with $Pr^{3+}$ compared to doping with $Ce^{3+}$. Full or partial replacement of the RE-cation in the complex oxides does not lead to a significant decrease in the degree of photon-nPR. In most cases oxides with a mixed type of RE-cation, e.g. LYSO:Ce, LGSO:Ce or LuYAP:Ce show a higher degree of photon-nPR. The semiconductor scintillator ZnO did not show a photon-nPR significantly different from the photon-nPRs of other scintillators, mostly insulators. An important factor in the halides nonproportionality is the width of the valence band. As the width of the valence band increases from fluorides to chlorides, to bromides and to iodides, a higher value for the photon-nPR at 10 keV is observed.

ACKNOWLEDGMENT

We thank the scientists and technicians of the X-1 beamline at the Hamburger Synhrotronstrahlungslabor (HASY-LAB) synchrotron radiation facilities for their assistance. Also we want to express our gratitude to our colleagues Mikhail Alekhin, Francesco Quarati and Johan de Haas who have made a significant contribution to the experimental activities and data analysis. We would like to thank Alan Owens from the European Space Agency for sharing with us some of the beamtime at X-1, Vladimir Ouspenski from Saint-Gobain Crystals for the LYSO, LaBr$_3$, LaCl$_3$ and NaI crystals, Karl Kramer from University of Bern for SrI$_2$ crystals, Piotr Rodnyi from St. Petersburg State Politechnical University and Elena Gorokhova from the Scientific Research and Technological Institute of Optical Material Science for the ZnO ceramics, and Paul Schotanus from SCIONIX Holland B.V. for the CeBr$_3$ crystal.